\input harvmac.tex
\Title{\vbox{\baselineskip12pt\hbox{IFT-CSIC-99-7}
\hbox{hep-th/9903012}}}
{\vbox{\centerline{On The Heterotic Dipole At Strong Coupling}}}
\centerline{\bf Sudipta Mukherji}
\smallskip\centerline{\it Instituto de Fisica Te\'orica}
\smallskip\centerline{\it Universidad Aut\'onoma de Madrid, Madrid,
Spain}
\smallskip\centerline{e-mail: sudipta.mukherji@uam.es}
\vskip .3in

We analyse the dipole solution of heterotic string theory 
in four dimensions. It has the structure of monopole 
and anti-monopole connected by flux line (string).  
Due to growing coupling near the poles,
the length of the string diverges. However, exploiting the 
self-duality of heterotic string theory in four dimension,
we argue that this string is correctly described in terms of dual
variables. 
\Date{3/99}
\lref\schwarz{See for example J. Schwarz, {\it Lectures on Superstring
and M-theory dualities}, hep-th/9607201; M. Duff, R. Khuri
and J. Lu, Phys. Rep. {\bf 259} (1995) 213. }
\lref\gp{D. Gross and M. Perry, Nucl. Phys. {\bf B226} (1983) 29.}
\lref\ashoke{A. Sen, JHEP {\bf 9809} (1998) 023.}
\lref\sen{A. Sen, JHEP {\bf 9710} (1997) 002.}
\lref\davidson{A. Davidson and E. Gedalin, Phys. Lett.
{\bf B339} (1994) 304.}
\lref\macias{A. Macias and T. Matos, Class. Quant. Grav. {\bf
13} (1996) 345.}
\lref\sentwo{A. Sen, Nucl. Phys. {\bf B440} (1995) 421.}

\def \th {{\theta}}
\def \st {{\rm sin}^2 \th}

\def \p {r^2 - a^2 {\rm cos}^2\th }

\noindent {\bf 1. Introduction:}

There are various solitonic objects in superstring theory. 
Among those, most useful are the ones that are BPS saturated.
BPS saturated solitons preserve certain fraction of supersymmetry.
As a result, because of non-renormalisation theorems, certain
quantum numbers associated to these objects do not receive 
corrections as we increase the string coupling. It is because of
this key property, BPS saturated solutions in string theory
play an important role in testing various duality conjectures in 
string theory \refs{\schwarz}. 
However, in order to shed more light on dualities,
one has to come out of these BPS objects. One kind of
such non-BPS states have been in focus recently \refs{\ashoke}. They are 
stable (even though non-renormalisation theorems are not
applicable for these states) because they are the lowest states
of certain quantum numbers. There is nothing to which 
they can decay. 

On the other hand, in various works, the dualities of string theory
have been exploited in order to understand instabilities associated with
the non-BPS states. In \refs{\sen} for example, the focus was on 
a D6 and anti-D6 brane pair of IIA string theory in ten dimension.
This is an unstable object in IIA theory. When the separation
between them is of the order of string length, there is tachyonic
instability. This tachyonic instability, however, was argued 
to be absent in the strongly coupled version of IIA theory-the 
M theory. 

In this note, we will be interested in certain non-BPS
solitons in four dimensional heterotic string theory. They are
dipole like solutions somewhat similar in nature to that of 
Gross-Perry dipole solutions~\refs{\gp} ~reduced to four dimension
\refs{\davidson, \macias}.
The heterotic dipole that we will be analysing has a string 
like structure (may be interpreted as a flux line) joining the 
two poles. However, measured in the string-frame, the length of
this string diverges. Also we notice that, at the same time, the 
string coupling blows up near the poles as well. In  section 3,
we then analyse the solution from the dual frame. Heterotic
string theory , at the classical level, is conjectured to 
be self-dual in four dimension 
under $SL(2,R)$ transformation. Thus the $SL(2,R)$ transformed dipole will
continue to be a solution the heterotic string in four dimension.
Exploiting this property, we analyse the nature of the
dual-dipole. We find, among other things, that the flux line 
connecting the poles is finite in length.

\bigskip
\noindent {\bf 2. Heterotic Dipole:}

In this section we will discuss the  dipole solution in heterotic
string theory. The heterotic string in four dimension is described by the
action
\eqn\one{S = \int d^4x {\sqrt{-G}}e^{-\Phi}
[ R_G + G^{\mu\nu}\partial_\mu \Phi\partial_\nu \Phi
- {1\over {12}}H_{\mu\nu\rho}H^{\mu\nu\rho}
- F_{\mu\nu}F^{\mu\nu}].}
Here $G_{\mu\nu}$ is the metric in the string frame. $\Phi$ is the
dilaton, $H_{\mu\nu\rho}$ is the anti-symmetric tensor and $F_{\mu\nu}$
is the electromagnetic  field strength.

This action has a dipole solution as found in \refs{\davidson}.
The field configurations for the dipole is given by:
\eqn\four{\eqalign{dS^2 &= ({{\Delta + a^2 \st}\over {\p}})[
-dt^2 + {\Delta \st (\p )^2 \over{(\Delta + a^2 \st)^2}} d\phi^2 \cr   
&+ {(\p )^2\over{\Delta + (M^2 + a^2)\st }}({dr^2\over \Delta}
+  d\theta^2) ],}}
\eqn\five{\eqalign{&\Phi =   {\rm log}{\p\over {\Delta + a^2 \st}}\cr
& A_{\phi} = {4aMr\st \over {\Delta + a^2\st}},\cr
& \Delta = r^2 - 2Mr - a^2.}} 
Here the solution is written in the Einstein-frame. The Einstein-frame
metric $(g_{\mu\nu})$ is related to the string-frame metric as 
$g_{\mu\nu} = e^{-\Phi} G_{\mu\nu}$.
 
First of all, we notice that the metric is asymptotically
flat. The dilaton $\Phi$ also goes to a constant as $r$ becomes
large. Furthermore, the asymptotic behaviour
of the gauge field reveals the dipole nature of the solutions.
The dipole moment can be read off and it is $4aM$ for the above 
field configuration. 

To have further understanding of the metric,
we would now look at the region $\Delta = 0, \theta = 0$ and 
$\Delta = 0, \theta = \pi$ as they are the possible locations
of singularities. This can be done by blowing up the coordinates 
near the corresponding points. For example, near $\Delta = 0, \theta
= 0$, we define coordinates \refs{\sen}
\eqn\ncoor{
(r_0 - 2M)\st = \tilde \rho (1 - {\rm cos}\tilde\theta ),~~
2(r - r_0) =  \tilde \rho (1 + {\rm cos}\tilde\theta).}
Here, $r_0 = M + {\sqrt{M^2 + a^2}}$ is the root of $\Delta = 0$.
Near the region $\theta = 0$ (keeping $a {\rm sin}^2\theta$ 
and $r - r_0$ finite), the field
configurations reduce to: 
\eqn\newf{\eqalign{ &
dS^2 =   -(1 + {M\over {\tilde \rho}})^{-1}  
dt^2 + (1 + {M\over {\tilde \rho}})(d{\tilde\rho}^2 + {\tilde\rho}^2
d{\tilde\theta}^2 + {\tilde\rho}^2 {\rm sin}^2\tilde\theta d\phi^2),\cr
&\Phi  =   {\rm log}(1 + {M\over{\tilde\rho}}),~~ A_{\phi} = M (1 - {\rm
cos}\tilde\theta).}}
The Ricci scalar can easily be calculated for the 
metric and is given by
\eqn\ricci{
R = {M^2\over {2\tilde\rho {(\tilde\rho + M)^3}}}.}
Thus, near $\theta = 0$, the metric behaves as anti-monopole
in heterotic string theory.
On the other hand, we can blow up the coordinates near $r = r_0,
\theta = \pi$. A similar computation as before reveals the 
structure of a monopole near the region  $r = r_0,\theta = \pi$.

We will now analyse the heterotic dipole in string
frame. Field configurations in string frame can be read off from
\four~after making the necessary metric scaling.
\eqn\hetf{\eqalign{dS^2 &= 
-dt^2 + {\Delta \st (\p )^2 \over{(\Delta + a^2 \st)^2}} d\phi^2 \cr
&+ {(\p )^2\over{\Delta + (M^2 + a^2)\st }}({dr^2\over \Delta}
+  d\theta^2) ,}}
\eqn\hetc{\eqalign{&e^\Phi =  {\p\over {\Delta + a^2 \st}}\cr
& A_{\phi} = {4aMr\st \over {\Delta + a^2\st}}.}}
From above we see that the non-zero electro-magnetic field
strength components are
\eqn\fstre{
F_{r\phi} = -{4aM (r^2 + a^2 {\rm cos}^2\theta)\st
\over{(\Delta + a^2\st )^2}},~~~
F_{\theta\phi} = {4a M r \Delta {\rm sin}2\theta \over
{(\Delta + a^2\st )^2}}.}
The metric can again be analysed near $\theta = 0, \pi$ exactly as
before in terms of coordinates $(t, \tilde \rho, \tilde\theta, \phi)$.
Near $\theta = 0$ is takes the form:
\eqn\newf{\eqalign{ &  
dS^2 =   -dt^2 + 
(1 + {M\over {\tilde \rho}})^2(d{\tilde\rho}^2 + {\tilde\rho}^2
d{\tilde\theta}^2 + {\tilde\rho}^2 {\rm sin}^2\tilde\theta d\phi^2),\cr
&\Phi  =   {\rm log}(1 + {M\over{\tilde\rho}}),~~ A_{\phi} = M (1 - {\rm
cos}\tilde\theta).}}
The curvature scalars related to the metric near $\theta = 0$, for
example, are
\eqn\hetr{
R = {2M^2\over {(\tilde\rho + M )^4}},~~~
R_{\mu\nu}R^{\mu\nu} = {2M^2(3\tilde\rho^2 + 2\tilde\rho M + M^2)
\over {(\tilde\rho + M )^8}}.}
Thus we notice that near $\tilde\rho = 0$ 
the curvature components are finite. 
The heterotic string coupling $g$ in our case is given 
by $e^{\Phi\over 2}$. From \hetc ~we see that
for large $r$, the coupling $g$ goes to constant which
has been normalised to $1$ for our solution. 
However, as we see from \hetc,  when $\Delta$ vanishes, $g$
{\it diverges} at $\theta = 0, \pi$. Thus the solutions given above
do not make sense. This is because, one expects, for large $g$, the 
action \one~itself would receive non-negligible corrections.
We will discuss the consequences later.
We also notice that $\int_0^{2\pi} {\sqrt {G_{\phi\phi}}} d\phi$ goes 
to zero as $r \rightarrow r_0$. This means that at $r = r_0$, there
is actually a string joining $\theta = 0$ and $\theta = \pi$. The
metric of the string that follows from \hetf ~is
\eqn\stringm{
ds^2 = -dt^2 + {(r_0^2 - a^2 {\rm cos}^2\theta)^2
\over {(M^2 + a^2)\st}}d\theta^2.}
One can thus calculate the length of the string as
\eqn\length{
l = \int_0^\pi {\sqrt {G_{\theta\theta}}} d\theta = 
{1\over {\sqrt{M^2 + a^2}}} \int_0^\pi {{r_0^2 - a^2 
{\rm cos}^2\theta}\over {{\rm sin}\theta }} d\theta.}
This is clearly {\it infinity}. However, we have noticed 
earlier that the string coupling is divergent at 
both the ends of the string.  

One can calculate the flux associated to the string by
integrating $A_\phi$ around the coordinate $\phi$ which
is
\eqn\fl{
\kappa = \int_0^{2\pi} A_\phi d\phi = {8\pi Mr_0\over a}.}

\bigskip

\noindent{\bf 3. The $SL(2,R)$ Transformed Solutions}:

In the last section, we saw that the effective string connecting
the monopole, anti-monopole system diverged as a consequence of
string coupling becoming strong near the poles.
What we can do however is to take the above configuration and
slowly decrease the string coupling $g$ by $SL(2,R)$ transformation.
Since, heterotic string theory is self-dual in four-dimension,
$SL(2,R)$ transformed string-frame metric and other fields
would continue to be solution of the same theory. Furthermore,
since a particular choice of $SL(2,R)$ metric inverts the string
coupling, the new coupling would be finite and small every where.
We thus first discuss a special class of $SL(2,R)$ transformed 
solution labeled by a parameter $0\le \delta \le \infty$ (see 
\sentwo ~for the constructional detail ).
For $\delta = \infty$, we get the configuration where new string
coupling is the inverse of the earlier. In what follows, 
we first construct the $SL(2,R)$ transformed solutions and
then discuss their behaviour in the light of section 2.

To do this, we have to remember that the heterotic string
has an anti-symmetric  three rank tensor field $H_{\mu\nu\sigma}$
as shown explicitly in \one.
For our earlier discussion, we set that $H_{\mu\nu\sigma}$ to zero and
hence it does not appear explicitly in the solutions \four-\five.
However, a generic $SL(2,R)$ transformation will indeed excite this
field.

Given the solutions in \hetf-\hetc, we can easily generate 
a subclass of $SL(2,R)$ dual solutions. In the string frame,
various fields  of the new solutions are given by:
\eqn\strongf{\eqalign{dS^2 &=
 {(r^2 - a^2 {\rm cos}^2\theta)^2 + \delta^2 (\Delta + a^2\st   
)^2 \over { (1+\delta^2) (r^2 - a^2 {\rm
cos}^2\theta)^2}}
[-dt^2 + {\Delta \st (\p )^2 \over{(\Delta + a^2 \st)^2}} d\phi^2 \cr
&+ {(\p )^2\over{\Delta + (M^2 + a^2)\st }}({dr^2\over \Delta}
+  d\theta^2)] ,}}
The dilaton and axion fields are respectively
\eqn\dilax{\eqalign{
&e^{\Phi^\prime} = {(r^2 - a^2 {\rm cos}^2\theta)^2 + \delta^2 (\Delta +
a^2\st
)^2 \over { (1+\delta^2)(\Delta + a^2\st ) (r^2 - a^2 {\rm
cos}^2\theta)}},\cr
&\Psi^\prime = {\delta [ (r^2 - a^2 {\rm cos}^2\theta)^2 -
(\Delta + a^2\st )^2]\over {(r^2 - a^2 {\rm cos}^2\theta)^2 + \delta^2
(\Delta + a^2\st)^2}}.}}
Furthermore, the electro-magnetic field components can be 
calculated from the formula
\eqn\nfs{
F^\prime_{\mu\nu} = {1\over{\sqrt{1 + \delta^2}}}(
F_{\mu\nu} + \delta\tilde F_{\mu\nu}).}
Here $\tilde F_{\mu\nu}$ is defined by
\eqn\twotwo{
\tilde F_{\mu\nu} = {1\over 2} {\sqrt{-g}}g_{\mu\mu^\prime}
g_{\nu\nu^\prime} \epsilon^{\mu\nu\rho\sigma} F_{\rho\sigma}.}
As it is straight forward to compute various components 
of $F^\prime_{\mu\nu}$ using the above formula, we do not 
display them here. However, we notice that for generic values
of the $SL(2,R)$ parameter $\delta$, $F^\prime_{\mu\nu}$ will
have electric as well as magnetic components.
In \dilax, we have defined $\Psi$ through
\eqn\sidef{
{\sqrt {-g}}e^{2\Phi} \epsilon^{\mu\nu\rho\sigma}\partial_\sigma \Psi
= -g^{\mu\alpha}g^{\nu\beta}g^{\rho\gamma}H_{\alpha\beta\gamma}.}

In terms of the coordinates $(t, \tilde\rho, \tilde\theta, \phi)$,
\strongf ~takes the form
\eqn\tstr{dS^2 = -{(\tilde \rho + M)^2 + \delta^2 \tilde \rho^2
\over {(1 + \delta^2) (\tilde\rho + M)^2}}dt^2 
+ {{(\tilde \rho + M)^2 + \delta^2 \tilde \rho^2
\over {(1 + \delta^2){\tilde\rho}^2}}} (d\tilde\rho^2 +
\tilde\rho^2d\tilde\theta^2 + \tilde\rho^2 {\rm sin}^2 \tilde\theta
d\phi^2).} 
It is easy to check that all the curvature components associated to this
metric are finite.
We also notice that the  new string coupling $g^\prime = e^{\Phi^\prime}$
has many nice features.
Firstly, for any $\delta$,
$g$ goes to $1$ for large $r$. Secondly, for finite $\delta$,
at $\theta = 0$ and  $\pi$, $g^\prime$ is finite. Thirdly,
for $\delta = \infty$, $g^\prime = g^{-1}$. Hence, the new
coupling becomes very weak at $\theta = 0, \pi$. 

At $r = r_0$, as before, the proper length along 
$\phi$ direction goes to zero. Thus there is still a string
like structure ending at $\theta = 0$ and $\pi$. 
This string is now described by the metric
\eqn\newstm{
ds^2 = 
 {({r_0}^2 - a^2 {\rm cos}^2\theta)^2 + \delta^2 a^4{\rm sin}^4\theta
 \over { (1+\delta^2) ({r_0}^2 - a^2 {\rm
cos}^2\theta)^2}}
[-dt^2 + {(r_0^2 - a^2{\rm cos}^2\theta )^2\over{ (M^2 + a^2)\st }}
d\theta^2].}
One can as before calculate the length of the 
string 
\eqn\nlen{l |_{\delta \rightarrow \infty} = {2a^2\over{\sqrt{M^2 +
a^2}}}.}
In the limit ${M\over a} \rightarrow 0$, we have from \nlen
~$l = 2a$.

To conclude, we summarize what we have done in this note. 
In section 2, we have analysed the heterotic dipole in four
dimensions as it would be probed by a fundamental string.
We noticed that there was a string (flux line) connecting the 
the two poles of the soliton. However, due to the string coupling
becoming strong near the poles, the length of the string turns out
to be infinite. In section 3, we then analysed the dual version 
of the solution. This is constructed by exploiting the $SL(2,R)$
self-duality property of the heterotic string in four dimensions. 
In the dual description, the string coupling is  
well-behaved all along the string joining the monopole, anti-monopole 
pair. As a consequence, we found the length of this string 
to be finite.
\bigskip
\noindent{\bf Acknowledgement:} I would like to
thank C\'esar G\'omez, Bert Janssen and Shibaji Roy for useful
conversations.
The work is supported by Ministerio de Educaci\'on y
Cultura of Spain and also through the grant CICYT-AEN 97-1678.

\vfill\eject
\listrefs
\bye